\documentclass[superscriptaddress,aps,pra,twocolumn,showpacs,longbibliography]{revtex4-2}
\usepackage[utf8]{inputenc}
\usepackage{graphicx,amsmath,amsfonts,amssymb}
\usepackage{color}
\usepackage[colorlinks=true, allcolors={blue}]{hyperref}
\usepackage{graphicx}
\usepackage{epstopdf}
\usepackage{float}
\usepackage{subfigure}
\newcommand{\orcid}[1]{\href{https://orcid.org/#1}{\includegraphics[width=7pt]{orcid.png}}}

\begin{document}

\title{Generic Two-Mode Gaussian States as Quantum Sensors}

\author{Pritam Chattopadhyay }
\email{pritam.chattopadhyay@weizmann.ac.il}
\affiliation{Department of Chemical and Biological Physics \& AMOS, Weizmann Institute of Science, Rehovot 7610001, Israel}

\author{Saikat Sur}
\email{saikat.sur@weizmann.ac.il}
\affiliation{Department of Chemical and Biological Physics \& AMOS, Weizmann Institute of Science, Rehovot 7610001, Israel}
\affiliation{Quantum Information Group, The Institute of Mathematical Sciences,
HBNI, CIT Campus, Taramani, Chennai 600113, India}

\author{Jonas F. G. Santos}
\email{jonassantos@ufgd.edu.br}
\affiliation{Faculdade de Ci\^{e}ncias Exatas e Tecnologia, Universidade Federal da Grande Dourados, Caixa Postal 364, CEP 79804-970, Dourados, MS, Brazil}

\begin{abstract}
Gaussian quantum channels constitute a cornerstone of continuous-variable quantum information science, underpinning a wide array of protocols in quantum optics and quantum metrology. While the action of such channels on arbitrary states is well-characterized under full channel knowledge, we address the inverse problem—namely, the precise estimation of fundamental channel parameters, including the beam splitter transmissivity and the two-mode squeezing amplitude. Employing the quantum Fisher information (QFI) as a benchmark for metrological sensitivity, we demonstrate that the symmetry inherent in mode mixing critically governs the amplification of QFI, thereby enabling high-precision parameter estimation. In addition, we investigate quantum thermometry by estimating the average photon number of thermal states, revealing that the transmissivity parameter significantly modulates estimation precision. Our results underscore the metrological utility of two-mode Gaussian states and establish a robust framework for parameter inference in noisy and dynamically evolving quantum systems.
\vskip 0.5cm
\end{abstract}

\maketitle

\section{Introduction}
Quantum metrology investigates the ultimate precision limits for estimating physical parameters embedded within quantum systems~\cite{PhysRevA.98.012114,giovannetti2011advances,RevModPhys.90.035005,schnabel2010quantum,toth2014quantum,haase2016precision,zwierz2012ultimate,morelli2021bayesian,seveso2017quantum,PhysRevLett.133.120601,Pritam2024QST,mukhopadhyay2024current,degen2017quantum,montenegro2024review}. A key challenge arises when the parameter of interest does not correspond to a direct observable but must be inferred through an optimized measurement strategy~\cite{paris2009quantum,seveso2017quantum,RevModPhys.90.035005}. The quantum \textit{Cramér–Rao bound} (QCRB)~\cite{Liu2020-zo} sets a fundamental lower limit on the mean squared error of such estimators, determined by the quantum Fisher information (QFI). Maximizing the QFI~\cite{Liu2020-zo,PhysRevA.89.032128,li2013entanglement,meyer2021fisher,rath2021quantum,frowis2016detecting,vitale2024robust,zhang2025krylov} enables the possibility of improving the sensitivity, especially under finite measurement resources, making it central to quantum-enhanced sensing protocols.

This framework has catalyzed breakthroughs across various domains, including gravitational wave detection via interferometric observatories like LIGO and VIRGO~\cite{Schnabel2010-lx, danilishin2020advanced}, phonon-based detection in Bose-Einstein condensates~\cite{Boixo2009-yw, Ngo2021-iy}, precision measurements in magnetometry and gravimetry~\cite{Hou2020-pm, Qvarfort2018-dr,aiello2013composite,PRXQuantum.5.010311,PhysRevLett.117.138501,stray2022quantum}, as well as the investigation of \textit{non-Hermitian systems} as quantum probes \cite{PhysRevA.109.062611,PhysRevLett.131.160801, santos2024NH}. In agreement with theoretical development, experimental implementation of metrological protocols has been progressing, for instance, in trapped ions systems~\cite{Marciniak2022-ob,gilmore2021quantum,reiter2017dissipative}, superconducting qubits~\cite{PhysRevLett.125.117701,danilin2024quantum,beaulieu2025criticality}, and in photonic devices~\cite{Pirandola2018-jf,munoz2024photonic}.

Among continuous-variable systems, Gaussian states serve as a cornerstone in estimation due to their experimental accessibility and full characterization via first and second moments of the quadrature operators~\cite{RevModPhys.84.621,Gerry2012-zn,kang2021experimental}. These states include coherent, thermal, squeezed, and entangled states, which serve as essential resources in quantum optics and information processing~\cite{Wang2007-lv}. The Gaussian formalism facilitates efficient QFI calculations, circumventing the complexities of infinite-dimensional Hilbert spaces~\cite{Safranek2015-bb,PhysRevA.88.040102,santos2024improving}. In particular, two-mode Gaussian operations, viz., \textit{two-mode squeezing} and \textit{beam-splitter}, play a pivotal role in generating and manipulating quantum correlations. Two-mode squeezing arises from nonlinear media, with its strength governed by the squeezing parameter~\cite{Gerry2012-zn}, while beam-splitters describe mode-mixing and optical losses~\cite{Perarnau-Llobet2020-hp}. Estimating these parameters effectively characterizes \textit{interaction strength} and \textit{environmental decoherence}, offering a pathway to optimized control in practical quantum technologies. 

In this work, we investigate the estimation in a class of \textit{general} two-mode Gaussian states, focusing on parameters that characterize key physical processes such as optical loss and nonlinear interactions. In particular, we focus on the estimation of the \textit{two-mode squeezing parameter}, linked to the nonlinear susceptibility of the medium, and the \textit{beam-splitter angle}, which models Gaussian loss channels. These parameters play a central role in quantum information protocols that rely on entanglement and mode-mixing operations. We further consider an \textit{asymmetric scenario} where one mode represents the system of interest and the other mode serves as an ancillary probe. Within this configuration, we analyze the QFI associated with the estimation of the average thermal photon number of the system. We explore how correlations introduced by the two-mode squeezing and the beam-splitter operations influence the estimation precision.

This article is organized as follows: Section \ref{TF} reviews the necessary concepts and tools to deal with Gaussian states and operations, especially those related to two modes. We also introduce the quantum Fisher information to study the relevant parameters characterizing the two-mode operations. In section \ref{Res}, we present our results concerning the estimation of the two-mode squeezing and the beam-splitter parameter. To conclude our investigation, we proceed to examine the influence of two-mode Gaussian operations on the precision of estimating the average thermal photon number of a single mode. For this purpose, we assume one mode to be the system of interest, while the other one is considered to be the ancillary system. In this last scenario, the two-mode Gaussian operations parameter, as well as the ancillary, are assumed to be completely known. Finally, section \ref{Con} draws the conclusions and final remarks.

\begin{figure}
    \centering
    \includegraphics[width=0.5\textwidth]{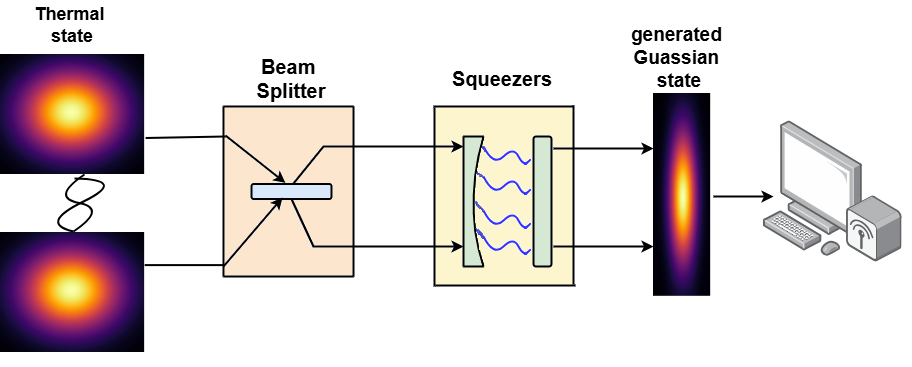}
    \caption{Schematic illustration of the generation of a two-mode Gaussian state using beam splitters and two-mode squeezers, followed by its sensing through measurement.}
    \label{fig:scheme}
\end{figure}

\section{Theoretical framework}\label{TF}

\begin{figure*}
  \begin{center}
       \subfigure[]{%
  \includegraphics[width=0.46\textwidth]{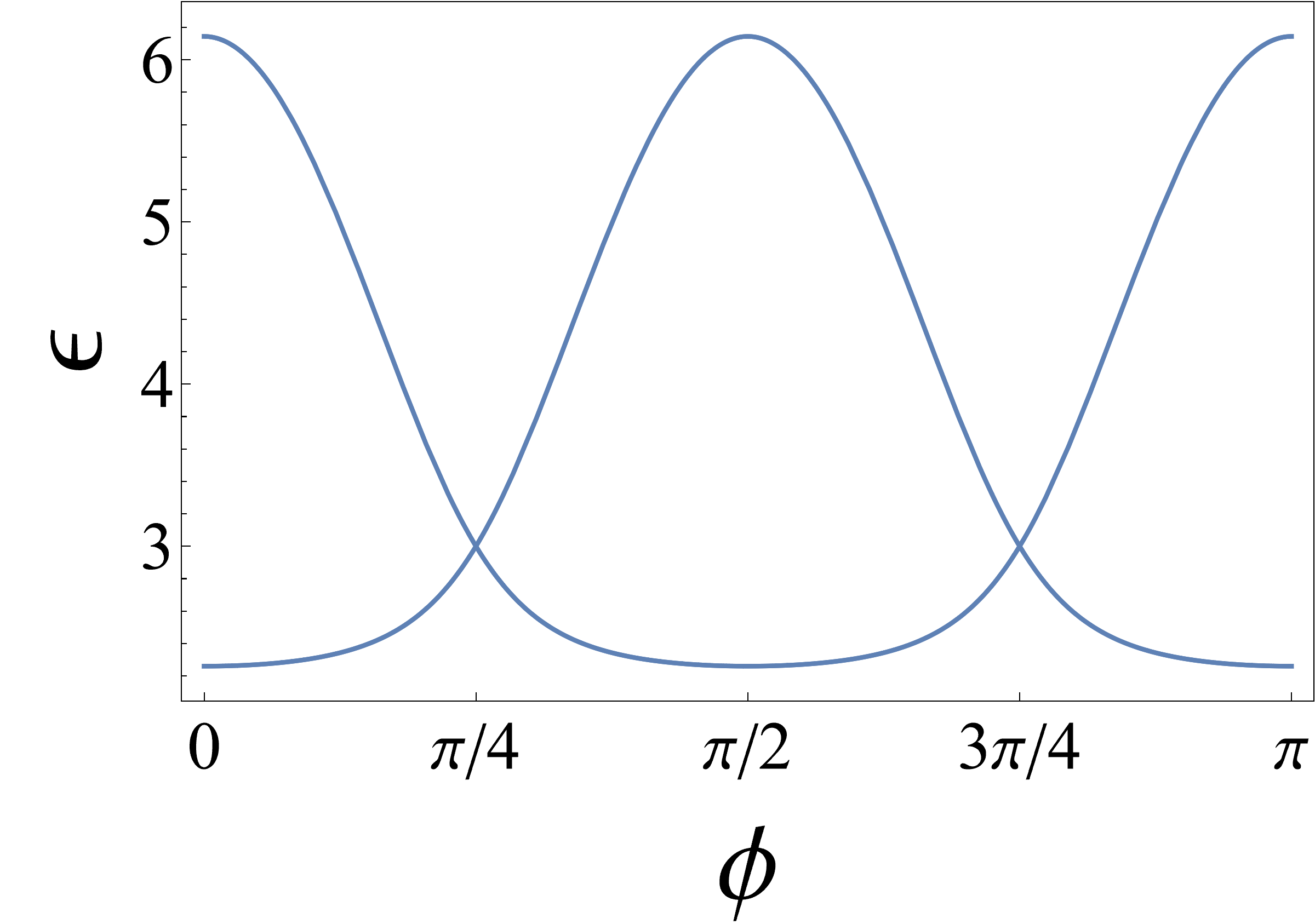}}
         \subfigure[]{%
  \includegraphics[width=0.48\textwidth]{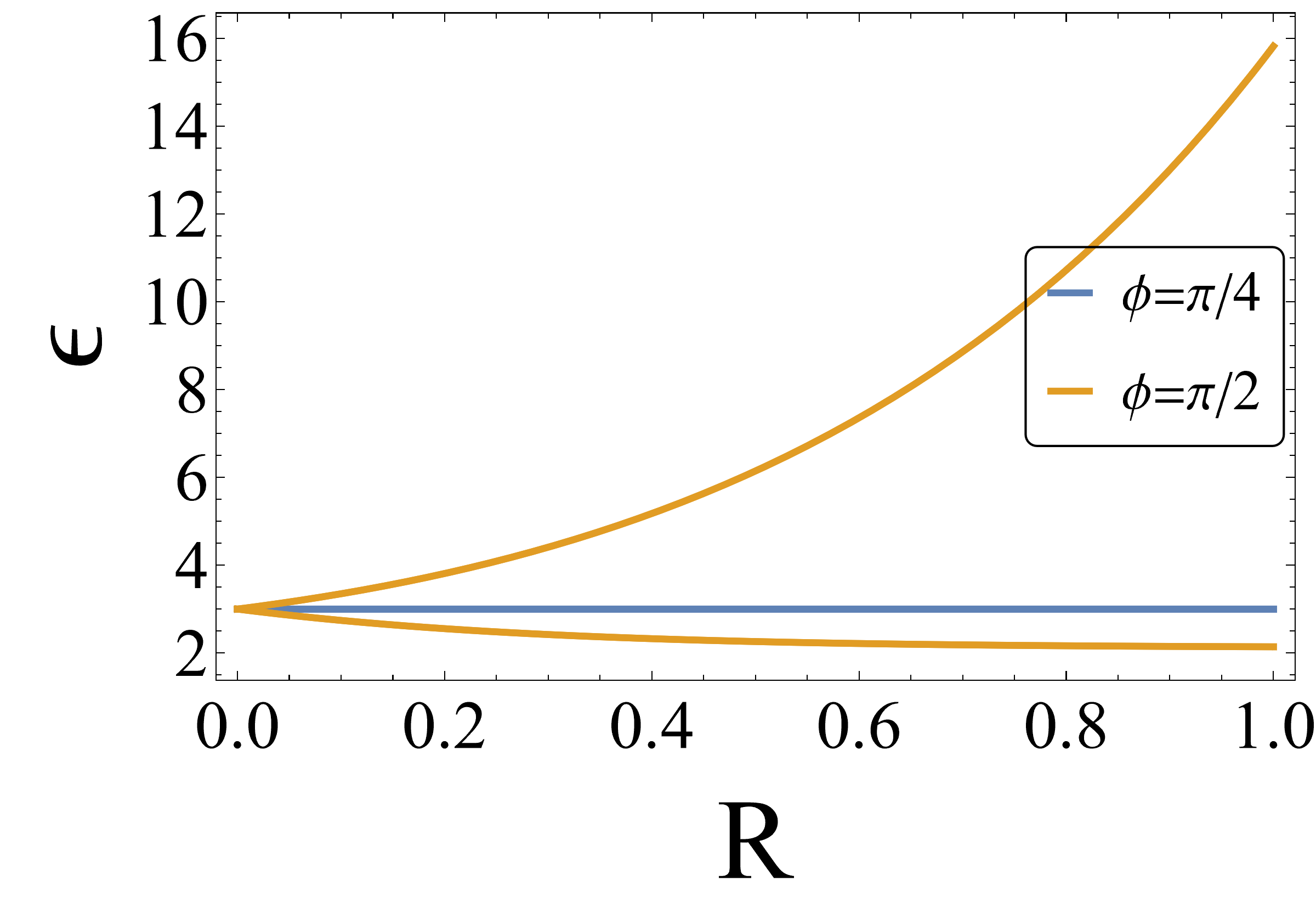}}
  \end{center}
\caption{(a) The spectrum of the covariance‐derived matrix $\mathcal{C}$ as a function of the beam‐splitter angle $\phi$, with the squeezing strength held at $R=0.5$,  exhibits two distinct level crossings. As $\phi$ sweeps from 
$0-\pi$, pairs of eigenvalues interchange their order precisely at 
$\phi=\pi/4$ and again at $\phi=3 \pi/4$. These sharp \textit{crossover} points signal resonant mixing between the two modes induced by the balanced beam splitter. (b) By contrast, if one fixes the beam‐splitter parameter at the special value  $\phi=\pi/4$ and instead varies the squeezing parameter $R$, the eigenvalues coalesce into a single non‐degenerate curve. In this configuration, the covariance matrix shows an enhanced symmetry, so that regardless of the squeezing amplitude, the entire spectrum remains perfectly degenerate. Whereas the degeneracy is lifted while we fix the beam-splitter parameter value at $\phi=\pi/2$. The average photon numbers for both cases are considered to be the same, i.e., $\bar{m}= \bar{n} =1$ for the thermal state pair.  }
      \label{fig:eigen1}
  \end{figure*}

\subsection{Gaussian states}
We begin by formalizing the framework for Gaussian states, a cornerstone in continuous-variable quantum information processing~\cite{RevModPhys.84.621}. An $N$-mode bosonic system, representing $N$ quantized electromagnetic field modes, can be modeled as a collection of $N$ quantum harmonic oscillators with an associated Hilbert space: $\mathcal{H}^{\otimes N} = \otimes_{k = 1}^N \mathcal{H}_k$.  Each mode $k$ is characterized by a pair of field operators,  $a_k$ and $a_k^\dagger$, representing the annihilation and creation operators, respectively. Furthermore, these operators are associated with the quadrature operators by the relations $q_k = a_k^\dagger + a_k$  and $p_k = i\left(a_k^\dagger - a_k\right)$. 

For the two-mode case ($N=2$), these operators can be compactly arranged into the field operator vector $\bold{b} = \left(a_1, a_1^\dagger,a_2, a_2^\dagger\right)^T$, satisfying the following set of commutation relations
\begin{subequations}
\begin{equation}
    \left[b_i , b_j\right] = \Omega_{ij},
\end{equation}
with $i, j = 1,...,4$ and $\bold{\Omega}$, with
\begin{equation}
    \Omega = \begin{pmatrix}
0 & \mathbf{I}_2 \\
-\mathbf{I}_2 & 0 
\end{pmatrix}.
\end{equation}
\end{subequations}
Equivalently, in terms of the quadrature operators we have $\bold{x} = \left(q_1,p_1,q_2,p_2\right)^T$, such that they must satisfy the relation $\left[x_i , x_j\right] = 2i\Omega_{ij}$, where we set $\hbar = 2$. 

For any two-mode bosonic system, a phase-space representation of a quantum state $\rho$ can be provided by the \textit{Wigner function} as~\cite{serafini2017quantum}
\begin{subequations}
\begin{equation}
    W\left(\bold{x}\right) = \int \frac{d^{4}\bold{\xi}}{\left(2\pi\right)^4} \exp\left[-i \bold{x}^T \bold{\Omega} \bold{\xi} Tr\left[\rho \exp\left[i \bold{x}^T \bold{\Omega} \bold{\xi} \right]\right]\right],
    \label{Wigner01}
\end{equation}
with $\bold{\xi} \in \mathbb{R}^2$ and the integral performed over all phase-space.
The Wigner function is normalized to unity, but it is generally a non-positive distribution. It is worth mentioning that, whenever $\rho$ is said to be Gaussian, this means that the Wigner function $W\left(\bold{x}\right)$ has a Gaussian distribution, implying that the first and second statistical moments provide the complete information to characterize the quantum state $\rho$. It must be stressed that any physical information described by an arbitrary parameter $\theta$ imprinted in the quantum state $\rho_\theta$ is fully accessible by the displacement vector $\bar{\bold{x}}_\theta$ and the covariance matrix $\sigma_\theta$ for a Gaussian state. 

The first moments (displacement vector) is defined by $\bar{\bold{x}} \equiv \text{Tr}(\bold{x} \rho)$, whereas the second moments (covariance matrix) is given by the elements $\sigma_{ij} = \langle x_i x_j + x_j x_i\rangle - 2\langle x_i\rangle \langle x_j\rangle$ in terms of the quadrature operators. The covariance matrix  $\bold{\sigma}$ must satisfy the uncertainty principle~\cite{serafini2017quantum}
\begin{equation}
    \sigma + i \bold{\Omega}\geq 0,
\end{equation}
which results in the positiveness of $\sigma$. Furthermore, for any Gaussian state, the Wigner function in Eq. (\ref{Wigner01}) is reduced to the following form
\begin{equation}
    W\left(\bold{x}\right) = \frac{\exp\left[-\left(1/2\right)\left(\bold{x} - \bar{\bold{x}}\right)^T \sigma^{-1} \left(\bold{x} -\bar{\bold{x}}\right)\right]}{\left(2\pi\right)^{4}\sqrt{\det \sigma}}.
\end{equation}
\end{subequations}

For a general two-mode Gaussian state, it is imperative to restrict our attention to Gaussian quantum operations---those transformations that map Gaussian states onto Gaussian states. A prominent class of such operations is the Gaussian unitaries, which are implemented via an evolution under a Hamiltonian of the form $U_\theta = \exp\left[-i H_\theta /2\right]$, where $H_\theta$ is a quadratic function of the canonical field operators. Based on this, the most general two-mode Gaussian state can be written as 
\begin{subequations}
\begin{equation}
\rho(\phi,R)=\mathcal{B}\left(\phi\right)\mathcal{S}\left(R\right)\rho_0 \mathcal{S}\left(R\right)^\dagger \mathcal{B}\left(\phi\right)^\dagger, 
    \label{generalstate}
\end{equation}
where $\mathcal{S}\left(R\right)$ and $\mathcal{B}\left(\phi\right)$ are the two-mode squeezing and the beam splitter operators, respectively. The former represents the \textit{pumping on a nonlinear crystal}, generating a pair of photons in two different modes, and is defined by the transformation  $U \rightarrow \mathcal{S}\left(R\right) = \exp\left[R\left(a_1 a_2 - a_1^\dagger a_2^\dagger\right)/2\right]$, while the latter constitutes a beam splitter transformation which is an example of an \textit{interferometer}, with the operator $U \rightarrow \mathcal{B}\left(\phi\right) = \exp\left[\phi\left(a_1^\dagger a_2 - a_1a_2^\dagger\right)\right]$.

The state described in Eq.~\eqref{generalstate} is fully characterized by two key parameters: the two-mode squeezing strength $R\geq 0$, and the beam-splitter mixing angle $0\leq \phi \leq \pi$. Two-mode squeezing and beam-splitter operations, representative of nonlinear and linear optical processes, respectively, constitute fundamental building blocks in the implementation of quantum information protocols involving two-mode bosonic systems~\cite{RevModPhys.77.513}. In addition, the input state  $\rho_0$ in Eq. (\ref{generalstate}), is assumed to be a tensor product of a pair of thermal states such that $\rho_0 = \rho_0^1\otimes\rho_0^2$, with
\begin{equation}
    \rho_0^i = \sum_{n = 0}^\infty \frac{\bar{n}^n_i}{\left(\bar{n}_i+1\right)^{n+1}}|n\rangle \langle n|,
\end{equation}
local thermal states, with average thermal numbers $\bar{n}_i$ and $\lbrace{|n\rangle\rbrace}_0^\infty$ is the Fock basis. In principle, one can choose the local states to be pure states in the Fock basis, such that $\rho_0 = |n_1\rangle \langle n_1 |\otimes |n_2\rangle \langle n_2 |$.

The transformations of the vector $\textbf{x}$ under the two-mode squeezing and beam-splitter operations are given respectively by the symplectic maps, $\textbf{x} \rightarrow \mathcal{\textbf{S}}\left(R\right)\textbf{x}$ and $\textbf{x} \rightarrow \mathcal{\textbf{B}}\left(\phi\right)\textbf{x}$, with 
\begin{equation}
    \mathcal{\textbf{S}}\left(R\right) = \left(\begin{array}{cc}
\cosh R\mathbb{I} & \sinh R\mathbb{Z}\\
\sinh R\mathbb{Z} & \cosh R\mathbb{I}
\end{array}\right),
\end{equation}
and
\begin{equation}
    \mathcal{\textbf{B}}\left(\phi\right) = \left(\begin{array}{cc}
\cos\phi\mathbb{I} & \sin\phi\mathbb{I}\\
-\sin\phi\mathbb{I} & \cos\phi\mathbb{I}
\end{array}\right),
\end{equation}
where $\mathbb{I}$ is the identity matrix and $\mathbb{Z} = \text{diag}\left(1,-1\right)$, both of order two.

The beam-splitter operation is commonly characterized by the parametrization  $\tau = \cos^2\phi$, where $\tau$ denotes the transmissivity and $\phi$ is the mixing angle. This unitary transformation plays a pivotal role in modeling Gaussian channels, particularly loss channels, within two-mode bosonic systems. Specifically, coupling one mode of the system to an ancillary vacuum mode via a beam splitter can effectively simulate dissipation or loss, allowing the beam splitter angle $\phi$ to serve as a proxy for characterizing the strength of the loss \cite{Perarnau-Llobet2020-hp}. Therefore, precise estimation of $\phi$ directly corresponds to identifying the parameters of the loss channel.

In contrast, the two-mode squeezing operation is intrinsically linked to the generation of entanglement and correlations between modes. This transformation can be physically realized through a parametric interaction in a nonlinear optical medium, driven by a strong external pump field \cite{Gerry2012-zn}. By tuning the pump frequency $\omega_p$ such that $\omega_p = \omega_a + \omega_b$, where $\omega_a (\omega_b)$ are the natural frequencies of the interacting modes, the effective Hamiltonian in the interaction picture takes the form
\begin{equation}
    H_I = i\hbar\left(\eta^\ast a b - \eta a^\dagger b^\dagger\right),
\end{equation}
\end{subequations}
where $\eta \equiv \chi^{(2)}\gamma$, with $\chi^{(2)}$ the second order nonlinear susceptibility, $\gamma$ the pumping field amplitude, and  $a,b$ denotes the mode operators. The quantity $\chi^{(2)}$ is a property of the nonlinear medium, and it is directly linked to the two-mode squeezing parameter $R$. Then, the estimation of $R$ is equivalent to performing the estimation of the susceptibility of $\chi^{(2)}$.

\subsection{Parameter estimation} \label{PE}

The estimation of an arbitrary parameter $\theta$ is performed through $\mathcal{N}$ measurements, resulting in a set of outcomes  $\mathcal{Q}_i$ of some observable $Q$. The precision in sensitivity $\left(\delta \theta\right)^2$ ($(\delta \theta)^2= \langle [\theta_s - \theta]^2\rangle$, where $\theta_s$ depends entirely on measurement outcomes), is restricted by the\textit{ Cram\'er-Rao bound} read as
\begin{equation}
    \left(\delta \theta\right)^2 \geq \frac{1}{\mathcal{N} \mathcal{I\left(\rho_\theta\right)}},
    \label{QFI01}
\end{equation}
where $\mathcal{I\left(\rho_\theta\right)}$ is known as  the quantum Fisher information (QFI) for a \textit{single measurement}~\cite{PhysRevA.89.032128,li2013entanglement,meyer2021fisher}. For an unbiased estimator in the limit of a large number of measurements, the relation (\ref{QFI01}) represents the best achievable sensitivity limit for $\theta$. For a general Gaussian state composed of $N$ bosonic modes, the QFI can be derived using, for instance, the Kraus operator representation for quantum measurements~\cite{serafini2017quantum}. The QFI can be expressed using various distance metrics~\cite{Safranek2015-bb}. As an example, for two-mode bosonic systems, we adopt the derivation based on the Bures distance~\cite{bures1969extension}. The Bures distance is a metric that quantifies the distinguishability between two quantum states, 
$\rho_1$ and $\rho_2$, and is expressed in terms of the \textit{Uhlmann fidelity} $\mathcal{F} (\rho_1,\rho_2) = (\text{tr} \sqrt{\sqrt{\rho_1} \rho_2 \sqrt{\rho_1}})^2$ as $d_\text{BD}^2 = 2(1- \sqrt{\mathcal{F} (\rho_1,\rho_2)})$ ~\cite{uhlmann1976transition}. The quantum Fisher information, which quantifies the ability to distinguish two neighboring quantum states characterized with the parameters $\theta$  and ${\theta+d\theta}$, is defined as~\cite{hayashi2006quantum}:
\begin{equation}
    \mathcal{I} (\rho_\theta) = 8 \lim_{d\theta \rightarrow 0} \frac{1- \sqrt{\mathcal{F} (\rho(\theta),\rho (\theta+ d\theta))}}{d\theta^2}.
\end{equation}

The task of calculating the QFI reduces to expanding the fidelity around the parameter $\theta$. For Gaussian states, the density matrix is fully determined by the first and second moments and is represented as $\rho_1= (\bold{x}_1,\sigma_1)$ and $\rho_2= (\bold{x}_2,\sigma_2)$, where ${\bold{x}}_i$ and $\sigma_i$ represent the first and second moments of the $i$-th state, respectively. For a two-mode Gaussian state, the fidelity is~\cite{RevModPhys.84.621}:
\begin{widetext}
    \begin{equation} \label{Fidelity}
    \mathcal{F} (\rho_1,\rho_2) = \frac{4~\text{exp} (-(\bold{x}_1-\bold{x}_2)^T (\sigma_1+\sigma_2)^{-1} (\bold{x}_1-\bold{x}_2))}{\left(\sqrt{|\sigma_1+\Omega| |\sigma_2 + \Omega|} + \sqrt{|I+\Omega\sigma_1 \Omega \sigma_2|} \right)-\sqrt{\left(\sqrt{|I+\Omega\sigma_1 \Omega \sigma_2|} + \sqrt{|\sigma_1+\Omega| |\sigma_2 + \Omega|}\right)^2-|\sigma_1 +\sigma_2|}},
\end{equation}
\end{widetext}
where 
 $|.|$ denotes the determinant.
The QFI for the two-mode Gaussian states is~\cite{Safranek2015-bb}


\begin{eqnarray}\nonumber\label{Fisher1}
     \mathcal{I} (\rho_\theta) &= & 2\dot{\mathbf{x}}^\dagger \sigma^{-1} \dot{\mathbf{x}}+ \frac{1}{2(|\mathcal{C}|-1)} \left(|\mathcal{C}|~ \text{tr} \left[\left(\mathcal{C}^{-1} \dot{\mathcal{C}}\right)^2\right] \right)\\ \nonumber
    &+ & \frac{1}{2(|\mathcal{C}|-1)} \left(\sqrt{|I + \mathcal{C}^2|} ~\text{tr} \left[\left((I+\mathcal{C}^2)^{-1} \mathcal{\dot{C}} \right)^2 \right] \right)\\ 
    & + &\frac{\Lambda_1^2 - \Lambda_2^2}{2(|\mathcal{C}|-1)} \left(-\frac{\dot{\Lambda}_1^2}{\Lambda_1^4-1} + \frac{\dot{\Lambda}_2^2}{\Lambda_2^4-1} \right),
\end{eqnarray}
where $\mathbf{x}$ denotes the displacement, $\mathcal{C}:= i \Omega\sigma(\theta)$, and the symplectic eigenvalues ($\Lambda_i$) of the second moments are $\Lambda_{1,2} = \frac{1}{2} \sqrt{\text{tr} [\mathcal{C}^2] \pm \sqrt{(\text{tr} [\mathcal{C}^2])^2 - 16 |\mathcal{C}|}}$. The first term captures the contribution of the dynamics of the moments of the Gaussian state concerning the parameter being estimated. The other terms in \eqref{Fisher1} represent the dynamic dependence of the second moments on the sensing parameter (see App.~\ref{App.A} for the detailed analysis and its closed form in terms of the eigenvalues of the $\mathcal{C}$ matrix). Eq.~\eqref{Fisher1} is employed in metrological protocols not only for investigating the potential benefits of the superradiant phase transition in Rabi's model~\cite{garbe2020critical} but also to improve parameter sensing by utilizing correlated Gaussian wave packets~\cite{porto2024enhancing}.

\section{Results}\label{Res}
We pursue the estimation of each parameter independently, deliberately suppressing the influence of the other to isolate their distinct metrological signatures. Concurrent (multivariate) estimation of both the beam splitter and squeezing parameters offers no substantial physical insights beyond minor qualitative distinctions. Consequently, we adopt a sequential approach to parameter sensing, enabling a clearer delineation of each parameter's impact on the quantum Fisher information (QFI) landscape.  

\subsection{Sensing beam-splitter parameter}
\begin{figure*}
\begin{center}
    \subfigure[]{
\includegraphics[width=0.435\textwidth]{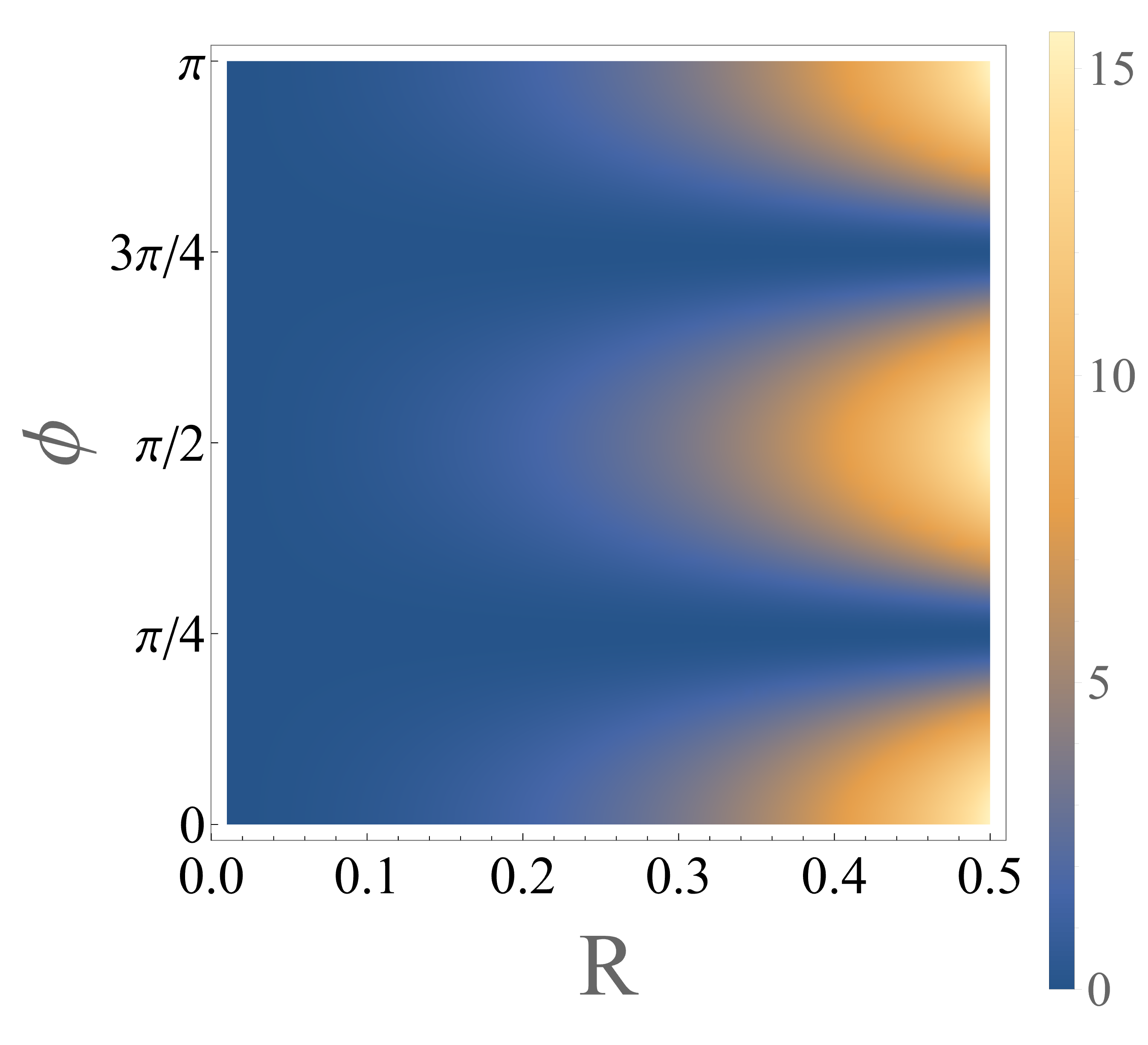} 
}
\subfigure[]{
\includegraphics[width=0.45\textwidth]{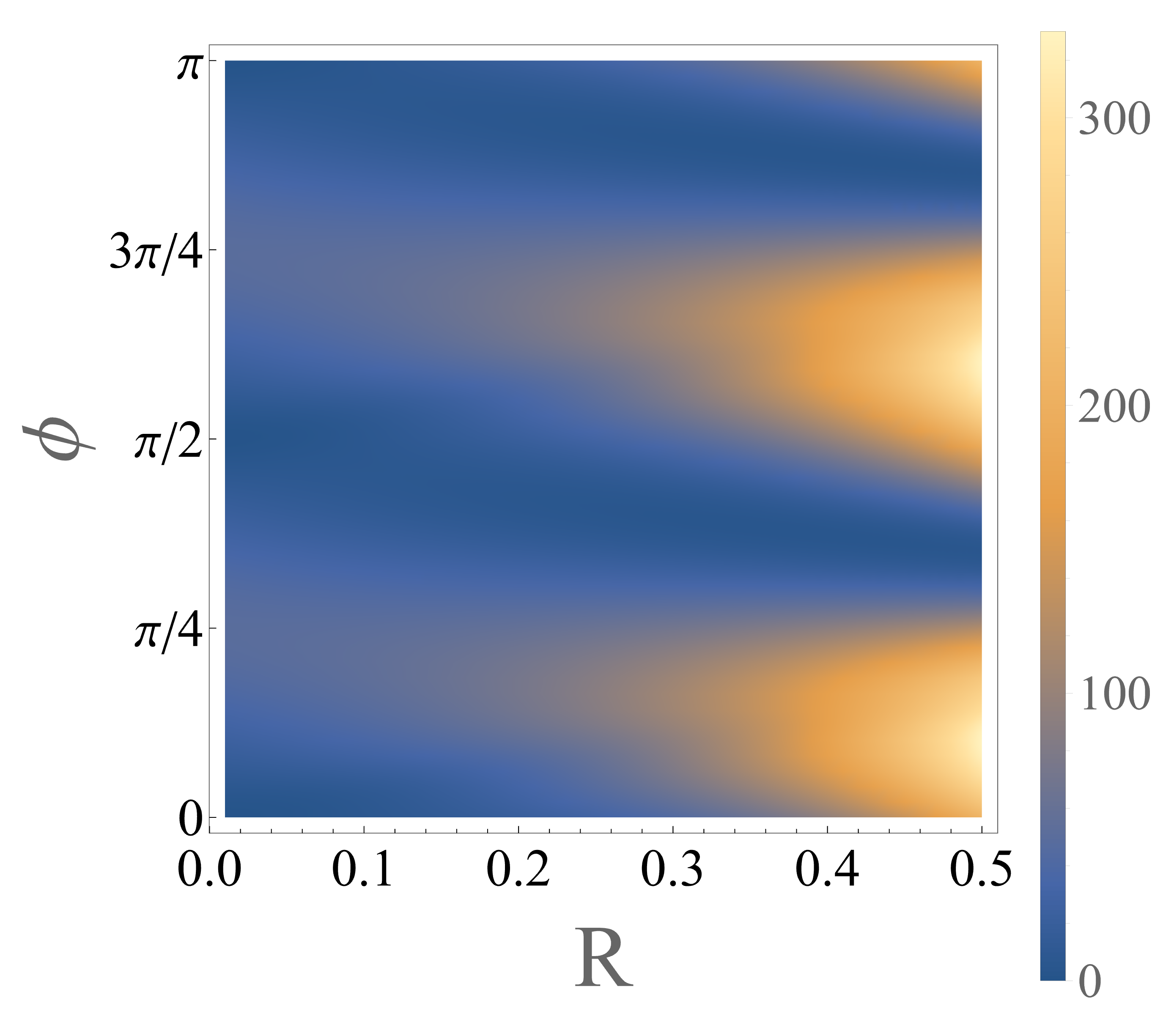} 
}
\end{center}
\caption{Quantum Fisher information associated with the estimation of the beam-splitter angle $\phi$. The analysis is performed for two distinct thermal regimes: (a) a balanced scenario with $\left(\bar{n}, \bar{m}\right) = \left(1, 1\right)$, and (b) a highly imbalanced configuration with $\left(\bar{n}, \bar{m}\right) = \left(1, 20\right)$. }
\label{QFI_phi}
\end{figure*}

\begin{figure*}
\begin{center}
    \subfigure[]{
\includegraphics[width=0.43\textwidth]{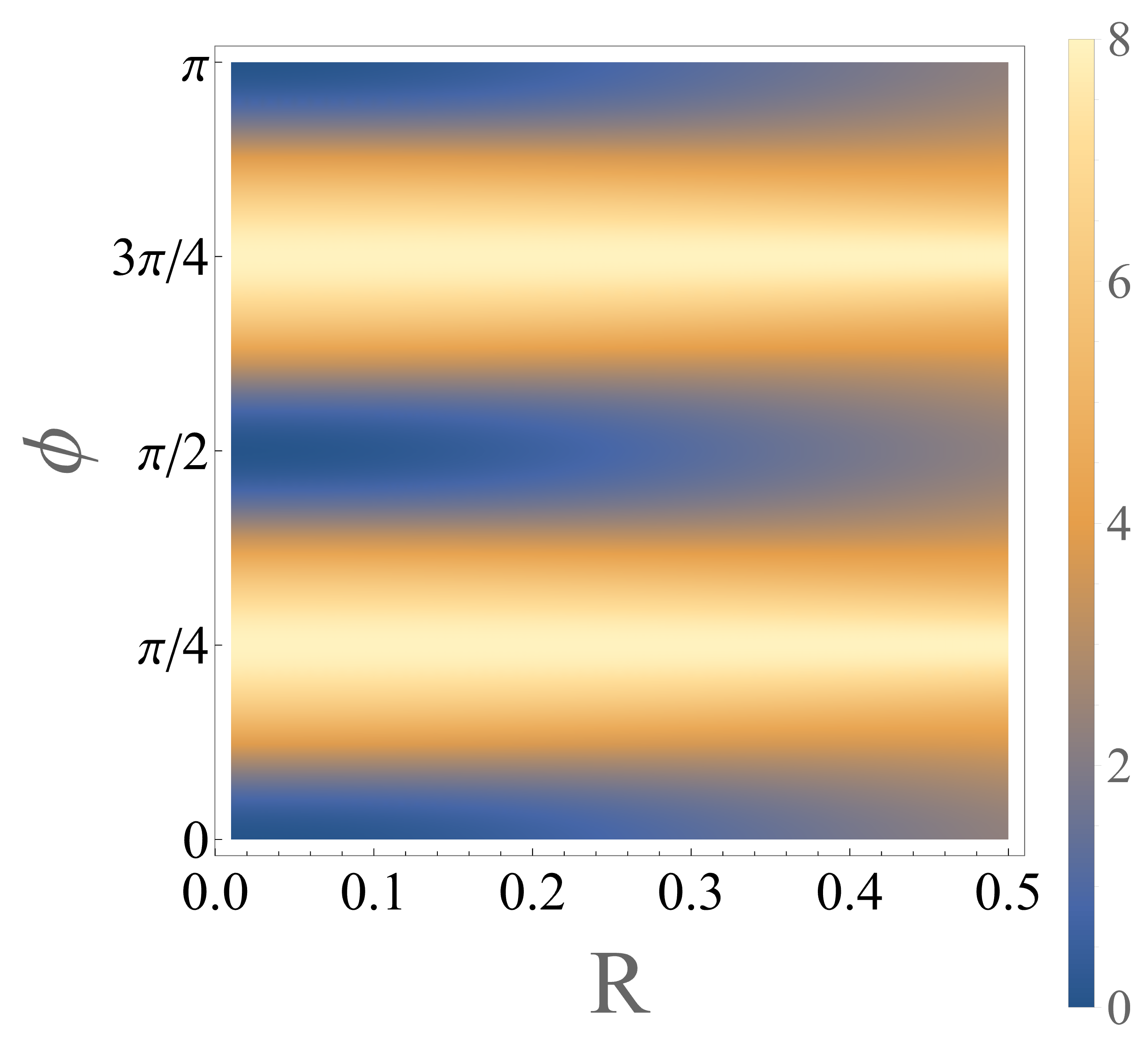} 
}
\subfigure[]{
\includegraphics[width=0.45\textwidth]{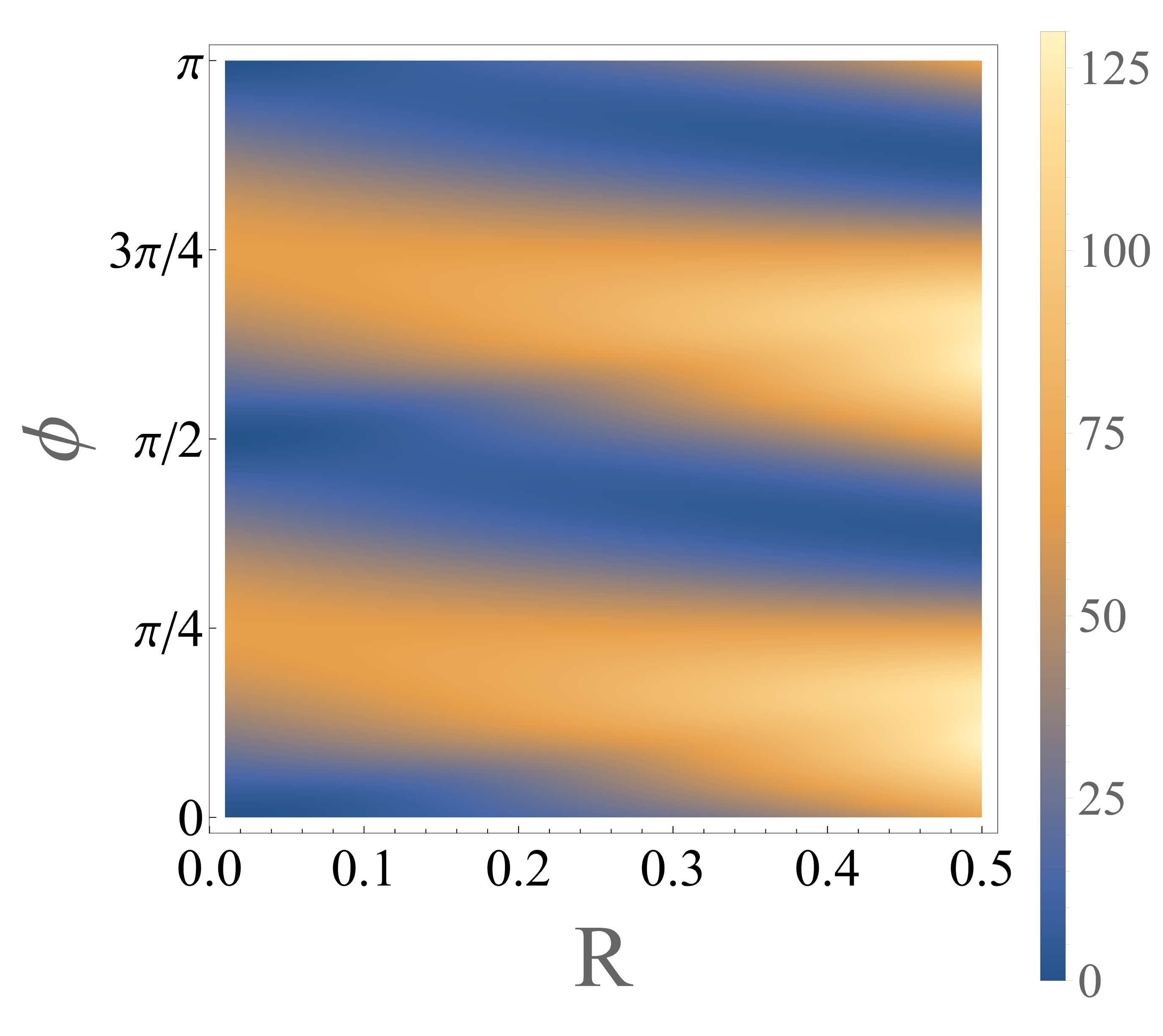}
}
\end{center}
\caption{Quantum Fisher information corresponding to the estimation of the two-mode squeezing parameter $R$. The evaluation is carried out for two thermal configurations: (a) a symmetric setup with $\left(\bar{n}, \bar{m}\right) = \left(1, 1\right)$, and (b) an asymmetric regime characterized by $\left(\bar{n}, \bar{m}\right) = \left(1, 20\right)$. }
\label{QFI_R}
\end{figure*}


We investigate the estimation of the beam-splitter parameter $\phi$ while constraining the squeezing parameter. The first statistical moments of the state are assumed to vanish, and the QFI is evaluated in a reduced form that depends solely on the symplectic eigenvalues of the covariance matrix (see App.~\ref{App.B}).

Figure~\ref{QFI_phi}-(a) depicts the QFI as a function of $\phi$ over the interval $[0,\pi]$ and $R$ over the interval $[0,0.5]$, revealing a pronounced peak near $\phi=\pi/2$. For this analysis, we assume that both modes initially possess equal thermal populations. At $\phi=\pi/4$, the beam splitter achieves optimal mixing of the two modes (Fig.~\ref{fig:eigen1}-(a)), resulting in a symmetrized output state. Under such symmetry and energy balance, the symplectic spectrum of the covariance matrix may become degenerate, diminishing sensitivity in certain parameter directions and consequently suppressing the QFI.

Conversely, at $\phi=\pi/2$, the beam splitter effectively performs a complete mode exchange, introducing maximal asymmetry into the system. This operation leads to a widening of the gap between the symplectic eigenvalues, which serves as a signature of enhanced distinguishability between neighboring quantum states. The emergence of a substantial eigenvalue gap indicates that the system is highly sensitive to infinitesimal variations in $\phi$, thereby elevating the QFI. Thus, the QFI is amplified in regimes where the state’s evolution in parameter space becomes sharply non-isotropic due to mode-asymmetry and correlation structure.

The thermal input states when exhibit an imbalance in photon number distribution, the spectral degeneracy is lifted, and the overall symmetry of the system is broken. This asymmetry leads to an increased rate of variation in the symplectic eigenvalue gap, thereby enhancing the sensitivity of the state to parameter changes. As a consequence, the QFI is significantly amplified with increasing photon number disparity, as illustrated in Fig.~\ref{QFI_phi}-(b).  



In the asymptotic limit $R\rightarrow 0$, an especially instructive scenario arises when $\Delta = 0$ (App.~\ref{App.B}), i.e., when both input modes possess the same average thermal occupancy. In this symmetric case, the eigenvalues of $\mathcal{C}$ become independent of $\phi$. This independence directly implies that the QFI, which is sensitive to how the eigenvalues vary with the parameter $\phi$, becomes insensitive to changes in $\phi$. In other words, the QFI with respect to $\phi$ vanishes, indicating that 
$\phi$ is unestimable from the state in this symmetric, unsqueezed configuration.

A key insight emerges at $R\gg 1$ when considering the symmetric thermal case, i.e., 
$\Delta=0$. In this scenario, although the dependence on 
$\phi$ becomes less pronounced due to the cancellation of asymmetric thermal terms, the eigenvalues of $\mathcal{C}$ continue to exhibit exponential sensitivity to the squeezing parameter $R$. Consequently, the QFI increases substantially with growing $R$, even in the absence of thermal imbalance. This underscores the central role of squeezing in enhancing quantum metrological precision, particularly when probing the parameter $R$ itself or estimating parameters indirectly affected by it.


\subsection{Sensing squeezing parameter}
In this section, we focus on the precision estimation of the squeezing parameter $R$ while constraining the beam-splitter parameter. Intriguingly, the behavior of the QFI for the estimation of $R$ reveals an inverse trend when compared to that of $\phi$ (Fig.~\ref{QFI_R}).

At $\phi=\pi/4$, where the beam splitter induces maximal mode mixing (Fig.~\ref{fig:eigen1}-(b)), the covariance matrix $\mathcal{C}$ exhibits a degenerate eigenvalue spectrum. It manifests the underlying symmetry in the system due to equal photon distribution across the thermal input modes. This spectral degeneracy, often associated with critical points of enhanced indistinguishability between modes, leads to a pronounced enhancement of the QFI (Fig.~\ref{QFI_R}-(a)), indicating optimal conditions for parameter estimation.

In contrast, at $\phi=\pi/2$, the symmetry of the system is broken, introducing maximal asymmetry in the energy distribution between the system and ancilla modes. This results in a pronounced widening of the gap between the symplectic eigenvalues, effectively reducing the sensitivity of the output state to variations in $R$. Consequently, the QFI is significantly suppressed in this regime. This contrasting behavior underscores the intricate interplay between modal correlations and metrological performance in Gaussian quantum sensing protocols.

Analogous to the behavior observed in the estimation of $\phi$, an imbalance in the photon number distribution of the thermal input states lifts the spectral degeneracy and disrupts the inherent symmetry of the system. This asymmetry amplifies the state’s susceptibility to variations in the parameter of interest, thereby significantly enhancing the QFI (Fig.~\ref{QFI_R}-(b)) and improving the precision of the estimation protocol.

For the limiting cases where $\phi =0, \pi/2$,  in particular, at $\Delta=0$ (App.~\ref{App.B}), the QFI associated with squeezing parameter estimation is expected to exhibit enhanced behavior since the system is fully symmetric and the squeezing manifests unimpeded by detuning noise. This makes the estimation of the squeezing parameter more precise, as the interference due to asymmetry is removed.

\begin{figure*}
\begin{center}
    \subfigure[]{
\includegraphics[width=0.45\textwidth]{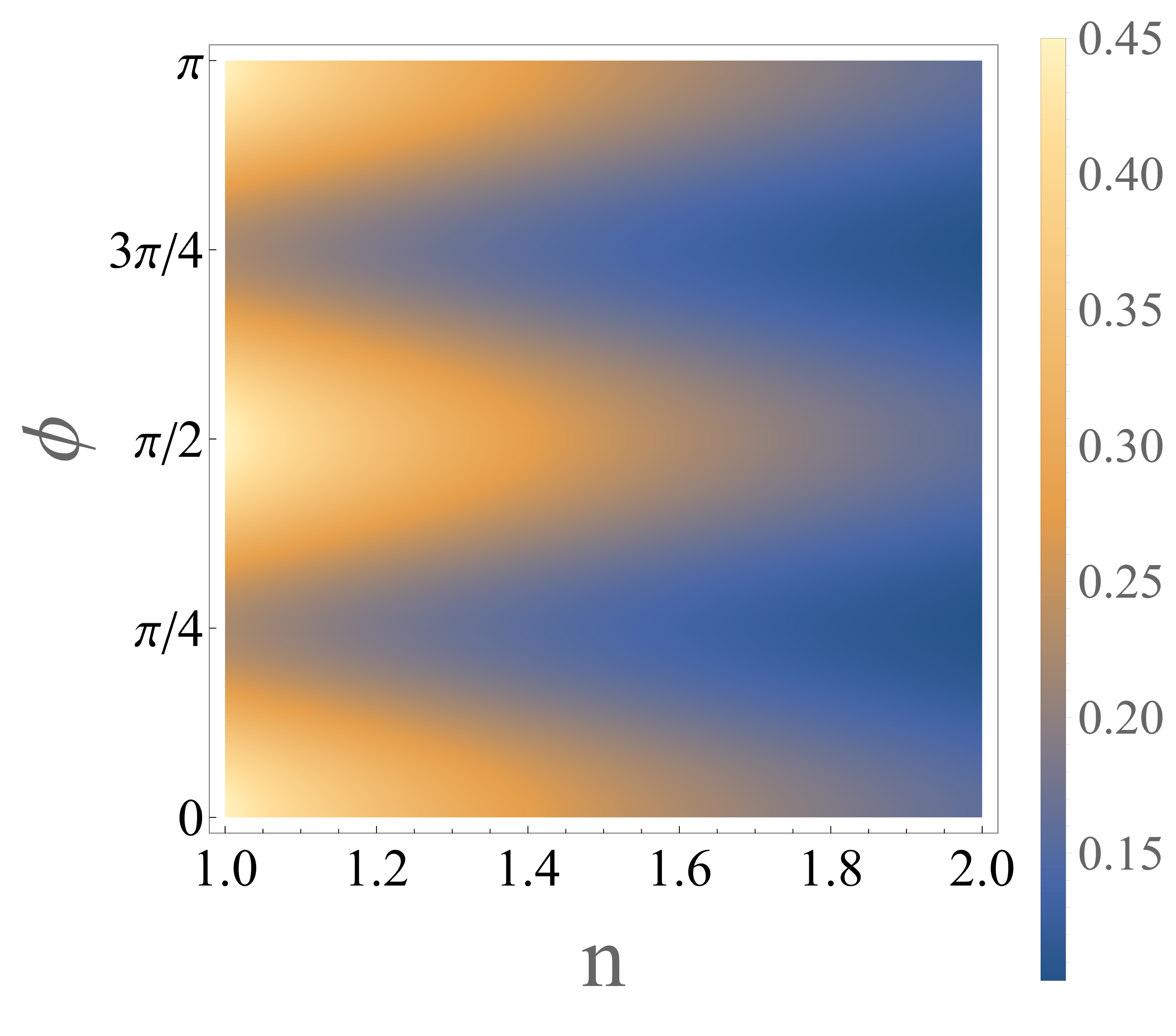} 
}
\subfigure[]{
\includegraphics[width=0.435\textwidth]{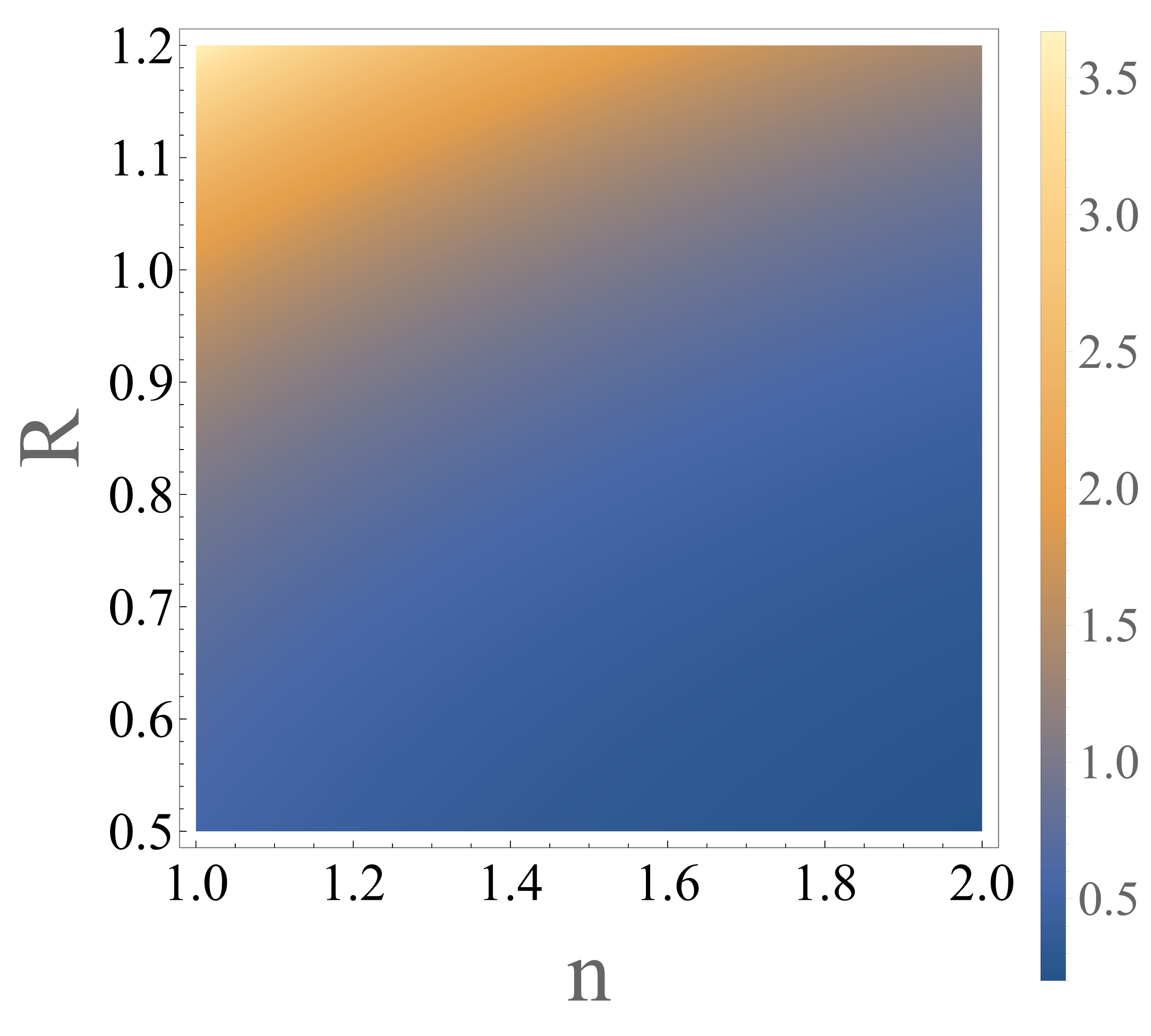} 
}
\end{center}
\caption{Quantum Fisher information for the estimation of the mean thermal number $\bar{n}$, considering the first photonic mode as the system and the second mode as an ancillary probe: (a) for $R=0$, (b) for $\phi=0$. We have also considered $\bar{m} = 1$ in both cases.}
\label{QFI_nbar}
\end{figure*}

\subsection{Effect of $\phi$ and $R$ on the estimation of the average thermal number}

Herein, we propose to exploit two-mode Gaussian unitaries, i.e., two-mode squeezing and beam-splitter operations, to perform quantum thermometry by estimating a single-mode mean thermal number. To this end, we designate the annihilation (creation) operators of the bosonic modes $a_1,(a_1^\dagger)$ and $a_2,(a_2^\dagger)$ to represent the system of interest and an ancillary mode, respectively, characterized by thermal occupation numbers $\bar{n}$ and $\bar{m}$. Within our framework, the parameters $\bar{m}$, $R$, and $\phi$ are assumed to be precisely known, and our objective is to accurately infer the unknown parameter $\bar{n}$. While we are here investigating the role played by two-mode unitaries on a single mode, the use of general single unitaries in quantum thermometry has been considered in Ref. \cite{PhysRevA.88.040102}. Figure~\ref{QFI_nbar} depicts the QFI $\mathcal{I}\left(\rho_{\bar{n}}\right)$ in two different scenarios, as a function of the beam-splitter parameter and changing $\bar{n}$, with $R = 0$ (Fig.\ref{QFI_nbar}-(a)), and as a function of the two-mode squeezing parameter and changing $\bar{n}$, with $\phi = 0$ (Fig.\ref{QFI_R}-(b)). For convenience, we have fixed $\bar{m} = 1$.

In Fig.~\ref{QFI_nbar}-(a), we show that, in terms of the transmissivity coefficient $\tau = \cos^2 \phi$, the QFI presents a maximum at $\tau = 0$ and $\tau = 1$, and this pattern repeats itself. On the other hand, the QFI is higher for small values of $\bar{n}$, decreasing as a function of it, as evidenced by previous works \cite{PhysRevResearch.5.043184,PhysRevA.110.052421}. Additionally, Fig.~\ref{QFI_nbar}-(b) illustrates the behavior of $\mathcal{I}\left(\rho_{\bar{n}}\right)$ in terms of the two-mode squeezing parameter $R$, where we can observe that there is a minimum value of $R$ in which the QFI admits an enhancement in the estimation. Thus, one can claim that having sufficient control over the parameters that involve two-mode Gaussian unitaries, it is possible to enhance the QFI, and then the uncertainty in $\bar{n}$ can be reduced.

\section{Conclusion}\label{Con}

Quantum metrology and sensing protocols have garnered significant attention in recent years, yielding both compelling theoretical advancements and experimental breakthroughs. In this work, we have focused on two-mode Gaussian states playing the role of quantum probes, and investigated the parameter estimation of general two-mode Gaussian operations, namely, the two-mode squeezing and the beam splitter operations. 

In our investigation, the precision estimation of both the beam splitter transmissivity and the squeezing parameter is found to be intrinsically governed by the symplectic eigenvalues of the covariance matrix. Specifically, while estimating the beam splitter parameter, a pronounced enhancement in the quantum Fisher information (QFI) emerges as a direct consequence of the widening spectral gap between the symplectic eigenvalues—an effect attributed to efficient intermodal energy exchange induced by the beam splitter operation. Conversely, in the case of estimating the squeezing parameter, the QFI attains a peak in regimes where optimal mixing of the two modes occurs, corresponding to conditions that symmetrically redistribute quantum fluctuations. This distinct behavior underscores the critical role played by the spectral structure of Gaussian states in quantum metrology, revealing that tailoring the interplay between modal interactions and state symmetries can significantly boost estimation sensitivity.

Finally,  we explored the influence of the squeezing parameter $R$ and the beam splitter parameter $\phi$  on the precision of estimating the local average thermal number. To this end, we proposed the configuration wherein one mode played the role of the system of interest, while the second mode was considered the ancillary system. Our analysis reveals that precise control over the interacting parameters significantly amplifies the QFI, thereby enabling a substantial reduction in the estimation uncertainty of the thermal occupation number. We anticipate that these findings will pave the way for further theoretical developments and experimental realizations in high-precision parameter estimation within continuous-variable quantum systems.\\

\section*{Acknowledgment}
J. F. G. Santos acknowledges CNPq Grant No. 420549/2023-4, Fundect Grant No. 83/026.973/2024, and Universidade Federal da Grande Dourados for support. P.C. acknowledges the support from the International postdoctoral fellowship from the Ben May Center for Theory and Computation.

\bibliography{Refs}

\newpage
\onecolumngrid

\appendix

\newpage

\section{Derivation of fisher information for two-mode Gaussian state\label{App.A}}
The Quantum Fisher Information (QFI) serves as a pivotal metric that quantifies the statistical distinguishability between infinitesimally close quantum states parameterized by a given variable. It encapsulates the ultimate sensitivity of a quantum state with respect to variations in that parameter, thereby establishing the fundamental precision bound in quantum estimation theory. QFI for the parameter $\theta$ is 
\begin{equation}\label{B1}
\mathcal{I} (\rho_\theta) = 8 \lim_{d\theta \rightarrow 0} \frac{1- \sqrt{\mathcal{F} (\rho(\theta),\rho (\theta+ d\theta))}}{d\theta^2},    
\end{equation}
where $\mathcal{F}$ is the fidelity between two quantum states as defined in Eq.~\ref{Fidelity}. Consequently, the computation of the Quantum Fisher Information (QFI) is reduced to evaluating the second-order expansion of the quantum fidelity between neighboring states around the parameter value $\theta$, thereby capturing the local curvature of the statistical distance in parameter space.

Let us denote the Taylor expansion of an arbitrary matrix-valued function 
$\mathcal{A}(\theta)$ about a point $\theta$, up to second order in $d\theta$, as:
\begin{equation}
    \mathcal{A}(\theta + d\theta)=\mathcal{A}(\theta)+ \frac{d\mathcal{A}} {d\theta} d\theta+ \frac{1}{2} \frac{d^2\mathcal{A}} {d\theta^2} (d\theta)^2 + \mathcal{O}((d\theta)^3)
\end{equation}
To obtain a simplified expression for the Quantum Fisher Information (QFI) that involves only the first derivatives of the covariance matrix with respect to the parameter of interest, one can invoke Williamson’s theorem~\cite{williamson1936algebraic,simon1999congruences}, which states that any real, positive-definite covariance matrix $\sigma$ can be decomposed as $\sigma = SES^\dagger$, where $S$ is a symplectic matrix and $E= \oplus \nu_k$ is a diagonal matrix containing the symplectic eigenvalues. Substituting this decomposition into the fidelity-based expression (as given in Eq.~\ref{Fidelity}) for the QFI allows us to recast the QFI in a form that isolates the parametric dependence in terms of the symplectic spectrum and the derivatives of the symplectic transformation. The QFI is expressed as 
\begin{eqnarray}
  &&  \mathcal{I} (\theta) =\frac{1}{|E(\theta)|-1} \left(|E(\theta)| \left(\text{tr} (\Gamma_1)^2 -\text{tr} (\Omega \Gamma_1 E(\theta) \Omega \Gamma_1/E(\theta))\right) \right) 
    + \frac{1}{2} \text{tr} \left[\frac{\dot E(\theta)}{(E(\theta) +\Omega) E(\theta)} \right] + \dot{\bold{x}}^\dagger (\theta) \sigma^{-1} \dot{\bold{x}} (\theta) \nonumber \\
    && + \sqrt{1+E^2(\theta)} \left[\text{tr} \left(\frac{\Gamma_1}{1+E^2(\theta)}\right)^2 +  \text{tr} \left(\frac{E(\theta) \Omega \Gamma_1}{1+E^2(\theta)}\right)^2 - \text{tr} \left(\frac{\Gamma_1^2}{1+E^2(\theta)}\right)\right],
\end{eqnarray}
where $\Gamma_1 = S(\theta)^{-1} \dot S(\theta)$, and $\bold{x}$ denotes the displacement. In general, computing the Williamson decomposition of a generic covariance matrix $\sigma$ is nontrivial, particularly for high-dimensional or parameter-dependent Gaussian states. To circumvent this difficulty, an alternative formulation of the QFI has been derived, which bypasses the explicit construction of the symplectic diagonalization. The expression, given in Eq.~\eqref{Fisher1}, involves only the matrix $\mathcal{C}\equiv i \Omega \sigma$ (which is directly proportional to the covariance matrix $\sigma$), the displacement vector $\bold{d}$, and the symplectic eigenvalues of the state, thereby offering a more tractable route for analytical and numerical evaluations.

Let us now consider that the eigenvalues of $\mathcal{C}\equiv i \Omega \sigma$  are $\lambda_i$ where $i=1,2,3, 4$. The eigenvalues of this matrix are fundamental in characterizing the symplectic structure of the system and play a central role in determining various physical and metrological properties. The QFI can be elegantly expressed in terms of the eigenvalues of $\mathcal{C}$, encapsulating the intrinsic phase-space geometry of the system. The expression for the QFI with respect to one of the parameter, keeping the other parameters fixed, is 
\begin{equation}
    \mathcal{I}(\rho_\phi) = \frac{\prod_i \lambda_i}{2 (\prod_i \lambda_i -1)} \left(\sum_i \frac{\dot{\lambda}_i^2}{\lambda^2_i} \right)+ \frac{\prod_i \sqrt{\lambda_i^2 +1}}{2 (\prod_i \lambda_i -1)}  \left(\sum_i \frac{\dot{\lambda}_i^2}{(\lambda_i^2 +1)^2} \right) + \frac{\Lambda_1^2 - \Lambda_2^2}{2(\prod_i \lambda_i-1)} \left(-\frac{\dot{\Lambda}_1^2}{\Lambda_1^4-1} + \frac{\dot{\Lambda}_2^2}{\Lambda_2^4-1} \right),
\end{equation}
where $\dot{\lambda}_i$ denotes the derivative of the eigenvalue of $\mathcal{C}$ in terms of $\phi$. The symplectic eigenvalues in terms of the eigenvalues of $\mathcal{C}$ is
\begin{eqnarray}
    \Lambda_{1,2} = \frac{1}{2} \sqrt{\sum_i \lambda_i^2 \pm \sqrt{\sum_i \lambda_i^2 - 16 \prod_i \lambda_i}}.
\end{eqnarray}
This formulation reveals the rich interplay between the symplectic structure of the phase space and the quantum statistical distinguishability of neighboring states. The dependence of the QFI on the derivatives of $\lambda$ reflects how sensitively the eigenstructure of $\mathcal{C}$ responds to variations in one of the parameters, thereby quantifying the information content carried by infinitesimal changes in the parameter.

Such an expression is particularly valuable in the study of quantum metrology in Gaussian systems, as it allows for a tractable and geometrically insightful route to evaluating parameter estimation bounds, with potential applications ranging from quantum sensing to continuous-variable quantum information processing.

\section{Moments of the Gaussian states \label{App.B}}
A general two-mode Gaussian state can be constructed by applying a sequence of Gaussian unitary transformations to a thermal product state. The state $\rho$ is represented as
\begin{eqnarray}
\rho=\mathcal{B}\left(\phi\right)\mathcal{S}\left(R\right)\rho_0 \mathcal{S}\left(R\right)^\dagger \mathcal{B}\left(\phi\right)^\dagger, 
\end{eqnarray}
where $\left(R\right)$ and $\left(\phi\right)$ are the two-mode squeezing and the beam splitter operators, respectively. These operations map the initial state $\rho_0$ into a correlated Gaussian state, enabling the generation of mode-mixing in phase space.

The beam splitter transformation is represented in the quadrature basis by the symplectic matrix:
\begin{eqnarray}
    \mathcal{B}(\phi) = \begin{pmatrix}
   \cos(\phi) & 0 & \sin(\phi) & 0
 \\
     0 & \cos(\phi) & 0 & \sin(\phi)
 \\
    - \sin(\phi) & 0 & \cos(\phi) & 0
 \\
    0 & - \sin(\phi) & 0 & \cos(\phi)
\end{pmatrix}
\end{eqnarray}
This model's passive linear optical transformations conserve the total photon number.

The two-mode squeezing operation, which generates quantum correlations and entanglement between modes, governed by the squeezing parameter $R$, is represented by
\begin{eqnarray}
    \mathcal{S}(R) = \begin{pmatrix}
   \cosh(R) & 0 & \sinh(R) & 0
 \\
     0 & \cosh(R) & 0 & -\sinh(R)
 \\
     \sinh(R) & 0 & \cosh(R) & 0
 \\
    0 & - \sinh(R) & 0 & \cosh(R)
\end{pmatrix}
\end{eqnarray}
The state $\rho_0$ is a product of two thermal state $\rho_0 = \rho_0 (\bar{n}) \otimes \rho_0 (\bar{m})$, where 
\begin{equation}
     \rho_0 (\bar{n}) = \sum_{n = 0}^\infty \frac{\bar{n}^n}{\left(\bar{n}+1\right)^{n+1}}|n\rangle \langle n|.
\end{equation}
This thermal product state is diagonal in the Fock basis and lacks any initial coherence or correlations between modes.

To characterize the Gaussian state fully, we examine its first and second moments. The first moments, i.e., the expectation values of the quadrature operators, are given by $<a_i>$, where $a_i$ are the annihilation operators of the modes. By propagating through the sequence of Gaussian transformations and using the properties of thermal states (which have zero displacement), one can verify via straightforward algebra that all first moments vanish. The essential information about the Gaussian state is thus encoded in the covariance matrix $\sigma$. The components of the covariance matrix are
\begin{eqnarray}\label{eq:cov_matrix}  \nonumber
    \sigma_{11} & = & (-\bar{m} + \bar{n})\cos(2 \phi) +(1 + \bar{m} + \bar{n}) \cosh(2 R) +(1 + \bar{m} + \bar{n}) \sin(2 \phi) \sinh(2 R),\\ \nonumber
    \sigma_{12} & = & (\bar{m} - \bar{n}) \sin(2 \phi) + (1 + \bar{m} + \bar{n}) \cos(2 \phi) \sinh(2 R),\\ \nonumber
    \sigma_{13} & = & 0,\\ \nonumber
    \sigma_{14} & = & 0, \\ 
    \sigma_{22} & = & (\bar{m} - \bar{n})\cos(2 \phi) +(1 + \bar{m} + \bar{n}) \cosh(2 R)-(1 + \bar{m} + \bar{n}) \sin(2 \phi) \sinh(2 R), \\ \nonumber
    \sigma_{23} & = &  0, \\ \nonumber
    \sigma_{24} & = & 0, \\ \nonumber
   \sigma_{33} & = &(-\bar{m} + \bar{n}) \cos(2 \phi)+(1 + \bar{m} + \bar{n}) \cosh(2 R)  -(1 + \bar{m} + \bar{n}) \sin(2 \phi) \sinh(2R),  \\ \nonumber
   \sigma_{34} & = & 0,\\ \nonumber
   \sigma_{44} & = & (\bar{m} - \bar{n})\cos(2 \phi) +(1 + \bar{m} + \bar{n}) \cosh(2 R)+(1 + \bar{m} + \bar{n}) \sin(2 \phi) \sinh(2 R). \\ \nonumber
\end{eqnarray}
These expressions encapsulate the full structure of the covariance matrix of the transformed two-mode Gaussian state. The dependence on $\phi$ and $R$ reflects the influence of the beam splitter and squeezing operations, respectively, while the thermal parameters $\bar{n}$ and $\bar{m}$ govern the initial mixedness of the individual modes.

\subsection{Beam-splitter parameter Sensing for fixed $R$}

\begin{figure*}
  \begin{center}
       \subfigure[]{%
  \includegraphics[width=0.46\textwidth]{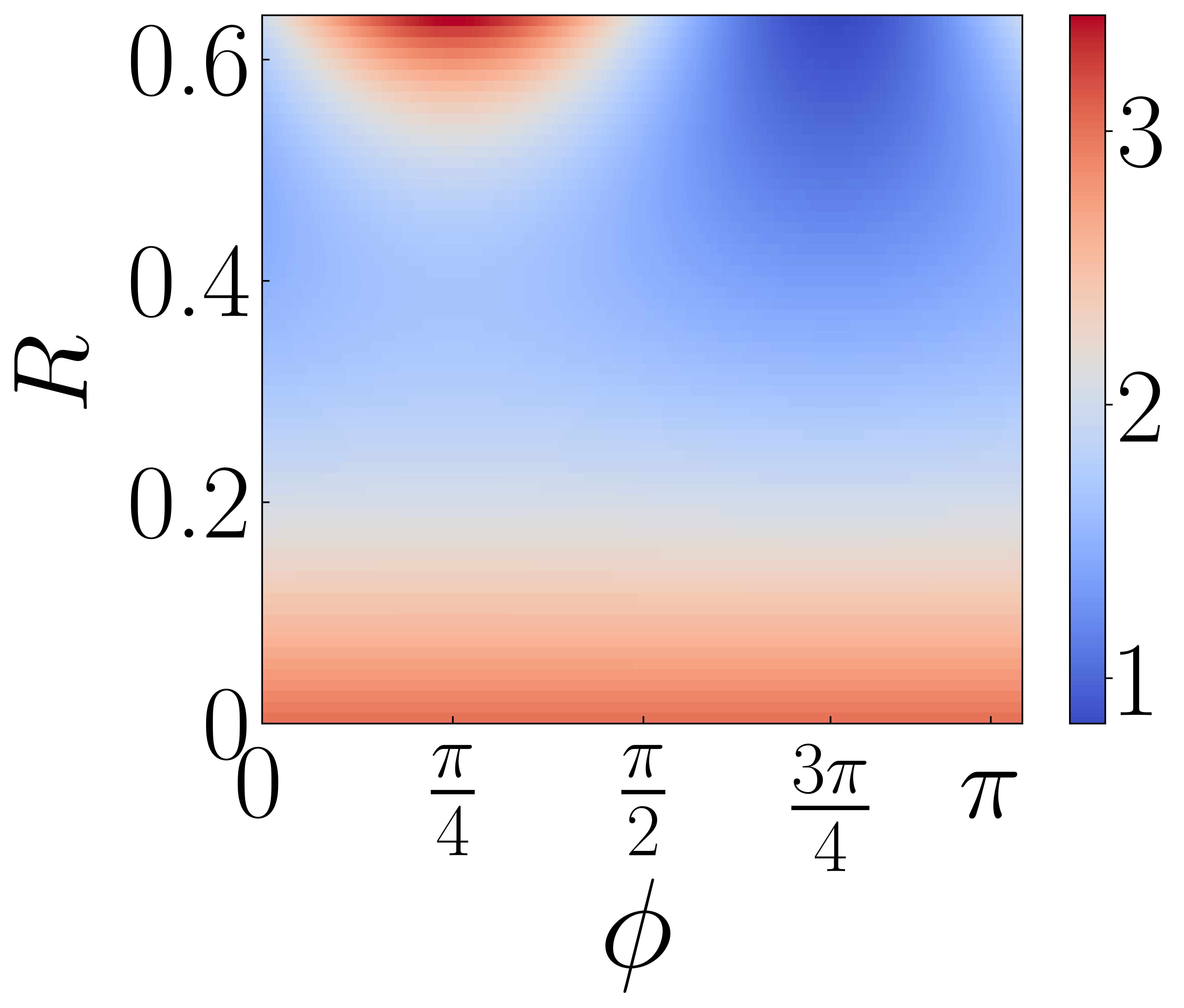}}
         \subfigure[]{%
  \includegraphics[width=0.48\textwidth]{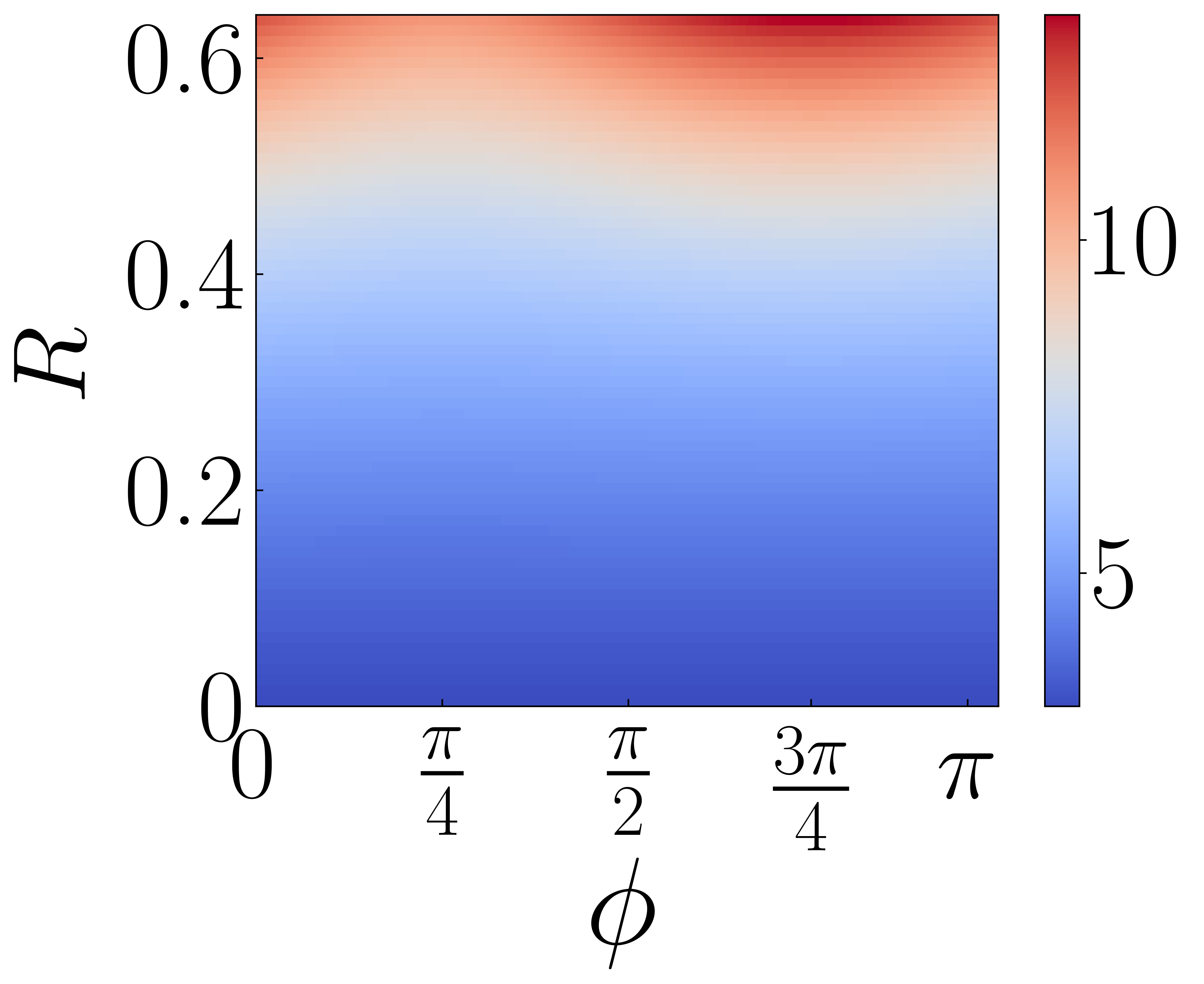}}
  \end{center}
      \caption{Eigenvalue spectrum of the $\mathcal{C}$ matrix in terms of the beam splitter and squeezing parameter for (a) $\Delta=0$ and (b) $\Delta=9$. }
      \label{fig:eigen}
  \end{figure*}

\textbf{Limiting case I $R=0$:} In the absence of two-mode squeezing, the Gaussian state undergoes only a passive linear transformation via the beam splitter. Under this condition, the covariance matrix simplifies significantly.  From \eqref{eq:cov_matrix}, the non-zero elements of the covariance matrix are expressed as 

\begin{eqnarray}\nonumber
    \sigma_{11} & = & -\Delta \cos(2 \phi) + (1 + S),\\ \nonumber
    \sigma_{12} & = &\sigma_{21}  = \Delta \sin(2 \phi), \\ \nonumber
    \sigma_{22} & = & \Delta \cos(2 \phi) + (1 + S),, \\ 
    \sigma_{33} & = & -\Delta \cos(2 \phi) + (1 + S),, \\ \nonumber
    \sigma_{44} & = & \Delta \cos(2 \phi) + (1 + S), \\ \nonumber
\end{eqnarray}
where $\Delta = \bar{m} -\bar{n}$ and $S = \bar{n} + \bar{m}$. The eigenvalues of $\mathcal{C} \equiv i \Omega \sigma$ becomes
\begin{eqnarray}\nonumber
&&\lambda = \{ -i \sqrt{\mathcal{A}-\mathcal{B}}, i \sqrt{\mathcal{A}-\mathcal{B}}, -i \sqrt{\mathcal{A}+\mathcal{B}},i \sqrt{\mathcal{A}+\mathcal{B}}  \}\\ 
\end{eqnarray}

where $\mathcal{A}= -\Delta ^2 \cos ^2(2 \phi )- (S+1)^2$, and \\$\mathcal{B} = \sqrt{4\Delta^2 (1+S)^2 \cos^2(2\phi) - \Delta^2\sin^2 (2\phi)\left((1+S)^2 - \Delta^2\cos^2 (2\phi)\right)}$.

These expressions highlight the intricate dependence of the symplectic spectrum on the beam splitter angle $\phi$, as well as on the thermal asymmetry $\Delta$ and the total thermal noise $S$.\\




\textbf{Limiting case II $R \gg 1$:} In the asymptotic regime of large two-mode squeezing, the hyperbolic functions simplify significantly as: $\sinh{2R} = \cosh{2R} \approx \exp{(2R)}/2 \ggg 1$. Under this condition, the covariance matrix elements are dominated by the exponential growth due to the squeezing operation. The non-zero components of the covariance matrix asymptotically take the form:
\begin{eqnarray}\nonumber
    \sigma_{11} & = & -\Delta \cos(2\phi) +(1 + S) (1+\sin(2\phi)) e^{2R}/2 \approx \frac{1}{2} e^{2R} (1 + S) (1+\sin(2\phi)) ,\\ \nonumber
        \sigma_{12} & = & \sigma_{21} = \Delta \sin(2\phi)+ (1 + S)\cos(2\phi) e^{2R}/2 \approx \frac{1}{2} e^{2R} (1 + S) \cos(2\phi),\\ 
        \sigma_{22} & = &\Delta \cos(2\phi) +(1 + S) (1-\sin(2\phi)) e^{2R}/2 \approx \frac{1}{2} e^{2R} (1 + S) (1-\sin(2\phi)),\\ \nonumber 
    \sigma_{33} & = & -\Delta \cos(2\phi) +(1 + S) (1-\sin(2\phi)) e^{2R}/2 \approx \frac{1}{2} e^{2R} (1 + S) (1-\sin(2\phi)), \nonumber\\
        \sigma_{44} & = & \Delta \cos(2\phi) +(1 + S) (1+\sin(2\phi)) e^{2R}/2 \approx \frac{1}{2} e^{2R} (1 + S) (1+\sin(2\phi)). \nonumber
\end{eqnarray}


The eigenvalues are:
\begin{eqnarray}
 \lambda = \{0,0, \pm  \frac{1}{\sqrt{2}} e^{2R} (1 + S) \cos(2\phi) \}   
\end{eqnarray}

These expressions illustrate the complex interplay between the squeezing parameter $R$, the thermal imbalance $\Delta$, and the beam-splitter mixing angle $\phi$. In particular, the exponential dependence on 
$R$ indicates that the squeezing transformation dominates the dynamics, with even minor variations in $R$ causing significant changes in the covariance structure and hence the eigenvalues of $\mathcal{C}$.



\subsection{Squeezing parameter Sensing for fixed $\phi$}

\textbf{Limiting case I $\phi=0$:}
In this regime, we analyze the behavior of the covariance matrix associated with the quantum state of interest when the beam splitter parameter is set to zero. 
The non-zero elements of the covariance matrix are
\begin{eqnarray}\label{phi0}  \nonumber
    \sigma_{11} & = &- \Delta+ (1 + S) \cosh(2 R),\\ \nonumber
     \sigma_{12} & = & (1 + S) \sinh(2 R),\\ \nonumber
    \sigma_{22} & = & \Delta + (1 +S) \cosh(2 R),\\ 
    \sigma_{33} & = &  -\Delta + (1 + S) \cosh(2 R), \\ \nonumber
   \sigma_{44} & = & \Delta + (1 + S) \cosh(2 R).
\end{eqnarray}
These expressions reflect the internal correlations of the quadrature components of the squeezed thermal state. The squeezing strength is encoded in the hyperbolic functions of $2R$, where $R$ denotes the squeezing parameter, while $S$ is associated with the thermal excitation or noise level, and 
$\Delta$ captures an asymmetry or detuning.


The eigenvalues of the $\mathcal{C}$ are
\begin{eqnarray}\nonumber \label{eqqnmnm}
&&\lambda = \{ -i \sqrt{\mathcal{F}-\mathcal{G}}, i \sqrt{\mathcal{F}-\mathcal{G}}, -i \sqrt{\mathcal{F}+\mathcal{G}},i\sqrt{\mathcal{F}+\mathcal{G}}  \}\\ 
\end{eqnarray}
where $\mathcal{F} = - \Delta ^2-  (1+S)^2 \cosh^2 (2R)$, and \\ $\mathcal{G} = (1+S)\sqrt{\cosh^2(2R) \sinh^2(2R) + 4 \Delta^2 \cosh^2(2R) - \Delta^2 \sinh^2(2R)}$.



The absence of $\Delta$ implies that the eigenfrequencies now depend solely on the squeezing $R$ and the thermal noise $S$, highlighting the fundamental role of quantum squeezing in governing the system’s evolution and its estimation capabilities.

\textbf{Limiting case II $\phi=\pi/2$:} In this regime, we analyze the behavior of the covariance matrix associated with the
quantum state of interest when the beam splitter parameter is set to $\pi/2$.  The non-zero elements of the covariance matrix are
\begin{eqnarray}\label{phi0}  \nonumber
    \sigma_{11} & = & \Delta+ (1 + S) \cosh(2 R),\\ \nonumber
     \sigma_{12} & = & -(1 + S) \sinh(2 R),\\ \nonumber
    \sigma_{22} & = & -\Delta + (1 +S) \cosh(2 R),\\ 
    \sigma_{33} & = &  \Delta + (1 + S) \cosh(2 R), \\ \nonumber
   \sigma_{44} & = & -\Delta + (1 + S) \cosh(2 R).
\end{eqnarray}



The eigenvalues of $\mathcal{C}$ are the same as Eq.~\eqref{eqqnmnm}.


\subsection{Thermal parameter sensing (thermometry)}
\textbf{Limiting case $\bar{m} \sim \bar{n}$:} In this parameter regime, we perform a detailed analysis of the structure and dynamics of the covariance matrix corresponding to the Gaussian quantum state under consideration. The non-zero elements of the covariance matrix are 
\begin{eqnarray}
    \sigma_{11} & = & (1 + 2\bar{m}) [\cosh(2 R)+ \sin(2 \phi) \sinh(2R)],\\ \nonumber
     \sigma_{12} & = & (1 + 2\bar{m}) \cos(2 \phi) \sinh(2 R),\\ \nonumber
    \sigma_{22} & = & (1 + 2\bar{m}) [\cosh(2 R)- \sin(2 \phi) \sinh(2R)],\\ 
    \sigma_{33} & = &  (1 + 2\bar{m}) [\cosh(2 R)- \sin(2 \phi) \sinh(2R)], \\ \nonumber
   \sigma_{44} & = & (1 + 2\bar{m}) [\cosh(2 R)+ \sin(2 \phi) \sinh(2R)].
\end{eqnarray}

The eigenvalues of $\mathcal{C}$ are
\begin{eqnarray}\nonumber 
&&\lambda = \{ -i \sqrt{\mathcal{J}-\mathcal{K}}, i \sqrt{\mathcal{J}-\mathcal{K}}, -i \sqrt{\mathcal{J}+\mathcal{K}},i \sqrt{\mathcal{J}+\mathcal{K}}  \}\\ 
\end{eqnarray}

where $\mathcal{J} = \sinh^2(2R)\sin^2(2\phi) - \cosh^2(2R)$, and $\mathcal{K} = \cos(2\phi) \sinh{2R}\sqrt{\cosh^2(2R) - \sinh^2(2R)\sin^2(2\phi)}$.  

The intricate dependence of the eigenvalues on $R$ and $\phi$ reveals the non-trivial structure of the quantum state. Particularly, the interplay between the squeezing amplitude $R$ and the phase $\phi$ governs the eigenvalue spectrum's spacing and curvature, which in turn critically influences the quantum Fisher information.

\end{document}